\documentclass[floatfix,amsmath,superscriptaddress,eqsecnum,amssymb,prb,twocolumn]{revtex4}
\usepackage{latexsym}
\usepackage{graphicx}

\newcommand{\be}{\begin{equation}}
\newcommand{\ee}{\end{equation}}
\newcommand{\ba}{\begin{eqnarray}}
\newcommand{\ea}{\end{eqnarray}}
\newcommand{\baa}{\begin{eqnarray*}}
\newcommand{\eaa}{\end{eqnarray*}}

\def\prl#1#2#3{Phys.\ Rev.\ Lett.\ {\bf #1}, #2 (#3)}

\def\prb#1#2#3{Phys.\ Rev.\ B {\bf #1}, #2 (#3)}

\def\rmp#1#2#3{Rev.\ Mod.\ Phys.\ {\bf #1}, #2 (#3)}

\def\be{\begin{equation}}
\def\ee{\end{equation}}
\def\ba{\begin{eqnarray}}
\def\ea{\end{eqnarray}}

\def\C60{A$_x$C$_{60}$}

\def\HgCu3{HgCa$_2$Cu$_3$O$_{8+y}$}
\def\HgCu4{HgBa$_2$Ca$_3$Cu$_4$O$_{10+y}$}
\def\TlCu{Tl$_2$Ba$_2$CuO$_{6+\delta}$}
\def\TlCu3{Tl$_2$Ba$_2$Ca$_2$Cu$_3$O$_{10+y}$}
\def\TlCu4{Tl$_2$Ba$_2$Ca$_3$Cu$_4$O$_{12+y}$}

\def\BiCu3{Bi$_2$Sr$_2$Ca$_{2}$Cu$_3$O$_y$}

\def\8LSCO{La$_{1.88}$Sr$_{.12}$CuO$_4$}
\def\110LNSCO{La$_{1.5}$Nd$_{0.4}$Sr$_{0.1}$CuO$_{4}$}
\def\stage4LCO{La$_{2}$CuO$_{4+\delta}$}
\def\Y248{YBa$_2$Cu$_4$O$_8$}

\def\NbSe2{NbSe$_2$}
\def\TaSe2{TaSe$_2$}
\def\TiSe2{TiSe$_2$}

\begin{document}

\title{Theory of Electron Nematic Order in LaOFeAs}
\author{Chen Fang}
\affiliation{Department of Physics, Purdue University, West
Lafayette, IN 47907}
\author{Hong Yao}
\affiliation{Department of Physics, Stanford University, Stanford, CA 94305}
\author{Wei-Feng Tsai}
\affiliation{Department of Physics, Stanford University, Stanford, CA 94305}
\affiliation{Department of Physics and Astronomy, University of California, Los Angeles, CA 90095}
\author{JiangPing Hu}
\affiliation{Department of Physics, Purdue University, West
Lafayette, IN 47907}
\author{Steven A. Kivelson}
\affiliation{Department of Physics, Stanford University, Stanford, CA 94305}

\begin{abstract}
We study a spin $S$ quantum Heisenberg model on the Fe lattice of the rare-earth oxypnictide superconductors.  Using both large $S$ and large $N$ methods, we show that this model exhibits a sequence of two phase transitions: from a high temperature symmetric phase to a narrow region of intermediate ``nematic'' phase, and then to a low temperature spin ordered phase.  Identifying phases by their broken symmetries, these phases correspond precisely to the sequence of structural (tetragonal to monoclinic) and magnetic transitions that have been recently revealed in neutron scattering studies of LaOFeAs.  The structural transition can thus be identified with the existence of incipient (``fluctuating'') magnetic order.
\end{abstract}
\date{\today}
\maketitle

\section{Context}
Of course, the big issue of the day is whether the physics of high temperature superconductivity\cite{Kamihara08, Ren, GFChen1, XHChen, GFChen2, HHWen} in the rare-earth oxypnictides is related to that in the cuprates.  In favor of this association is the observation that both are ``bad metals''\cite{bad}, and so presumably not well described by Fermi liquid theory in their normal states. Preliminary evidence\cite{Mu,Shan} suggests that the superconducting state in the oxypnictides, like that in the cuprates, has gapless nodal quasiparticle excitations, and hence, probably, an unconventional pairing symmetry.  Finally, there is tantalizing evidence that competing ordered states, and possibly %with a
an associated quantum critical point, %under or near the superconducting phase,
may play a role in both cases.\cite{mook,liu,GFChen2}

In the case of the cuprates, superconductivity is produced by doping a commensurate spin-ordered, insulating parent ``Neel'' state.  Probably Neel order does not coexist with superconductivity;  however, other ordered states, including spin and charge stripe (unidirectional density wave) ordered states,\cite{tranquada,rmp} an Ising nematic state\cite{ando,keimer,rmp} (about which, more later) and a form of time reversal symmetry breaking order\cite{bourges,kapitulnik} (whose character is still being debated)  seem to coexist (in some cases, at least) with superconductivity,  and possibly to vanish at quantum critical points somewhere under the superconducting dome.

The oxypnictides in question have chemical makeup
RO$_{1-x}$F$_x$FeAs, where R is a rare earth, and $x$ is the dopant
concentration;  the behavior (including the maximum superconducting
$T_c$) depends systematically on the particular choice of R.  The
situation with competing orders in the oxypnictides is only
beginning to be explored.  Undoped ROFeAs is not cleanly insulating,
but its resistivity is strikingly large ({\it e.g.} $\rho \sim $ 7
m$\Omega$-cm in SmOFeAs at room temperature\cite{liu,caveat}) for a metal;  it does, however, exhibit (in neutron scattering experiments on LaOFeAs\cite{mook,McGuire}) commensurate spin order below
$T_{SDW}$ = 135K.  Moreover, a closely associated structural
transition, which we wish to identify as the transition to an
``electron nematic phase'',\cite{nature} occurs at the slightly higher
temperature, $T_{\cal N}$ = 150K.\cite{mook}  The evolution of these orders
as a function of doping, $x$, has not yet been directly probed with
neutrons.  However, the sharp onset of an anomalous drop in the
resistivity occurs at $T_\rho=T_{\cal N}$ in the undoped $x=0$ material.
The onset temperature for the resistance drop has been tracked in
resistivity measurements for different $x$, and found to extrapolate to 0 at a
critical value of $x$ close to the point at which the superconducting
T$_c$ first reaches its maximum value Ð max[T$_c$] = 55K in
SmO$_{1-x}$F$_x$FeAs.\cite{liu} Assuming that the association
between T$_\rho$ and T$_{\cal N}$ persists, this means that there is an
electron nematic to isotropic quantum phase transition in the superconducting
dome in at least some members of both families of high temperature
superconductors, a suggestive evidence both of a common thread in the
behavior of both materials, and of the conjecture that nematic order
is, in some way, a crucial part of the physics.

\section{Introduction}
The purpose of this paper is to propose a unified perspective on the occurrence of both the magnetic and the structural phase transitions in undoped LaOFeAs.  To the extent that this proposal is correct, it justifies the identification of the observed structural transition with the occurrence of an ``electron nematic phase.''

Following the lead of two insightful recent papers by Yildirim\cite{yildirim} and by Si and Abrahams\cite{si}, we will treat undoped ROFeAs as if it were a magnetic (``Mott'') insulator, and hence we consider a simple model of localized spins on the iron sites interacting with neighboring spins by an antiferromagnetic superexchange interaction mediated through the intervening As atoms (Eq. \ref{H}, below).  Some microscopic justification for this approach is contained in those two earlier papers. In addition, the fact that the magnetic order\cite{mook} in ROFeAs is commensurate is, a priori, indicative of strong coupling physics.  Moreover, although the bare susceptibility
%, $\chi(\vec q,\omega)$,
 computed\cite{Mazin,Dong,sri} from the LDA band structure is peaked at the appropriate  ordering vector, ${\bf Q}$, the Fermi surface is not well nested and this peak is neither pronounced nor sharp.

However, it is important to acknowledge from the start that a model of localized spins  cannot be taken as a realistic representation of the electronic structure of ROFeAs.   The most obvious point is that ROFeAs is not an insulator!  At best, it is our hope that the magnetic and structural properties of this material can be qualitatively understood on the basis of the present model.  Even there, we will show that the small magnitude of the ordered moment, $m=0.35\mu_B$, at low temperature is inconsistent with the predictions of the present model.  Indeed, as pointed out by Si and Abrahams,  it is not even clear whether we should be considering a spin S=2 or S=1 model, in this context.  Still, the model is sufficiently simple that its behavior can be cleanly derived.  As we shall see, it inevitably exhibits two ordered phases (see Fig. \ref{fig:phase}) of precisely the character of those seen in experiment.  Moreover, numerous spectroscopic and a few thermodynamic predictions can be made on the basis of this model which are sufficiently robust that we may hope they transcend the deficiencies of the model. For instance, we predict the following: applying external pressure along the $z$ axis reduces the difference between the two transition temperatures $T_{\cal N}$ and $T_{SDW}$; there are sharp spin wave modes around ${\bf Q}^\prime=(0,\pi)$ (about which, see Section \ref {largeS}) with large spectral weight, which could be tested in future neutron scattering experiments.

 \section{The Model}
\begin{figure}
\includegraphics[width=9cm, height=6cm]{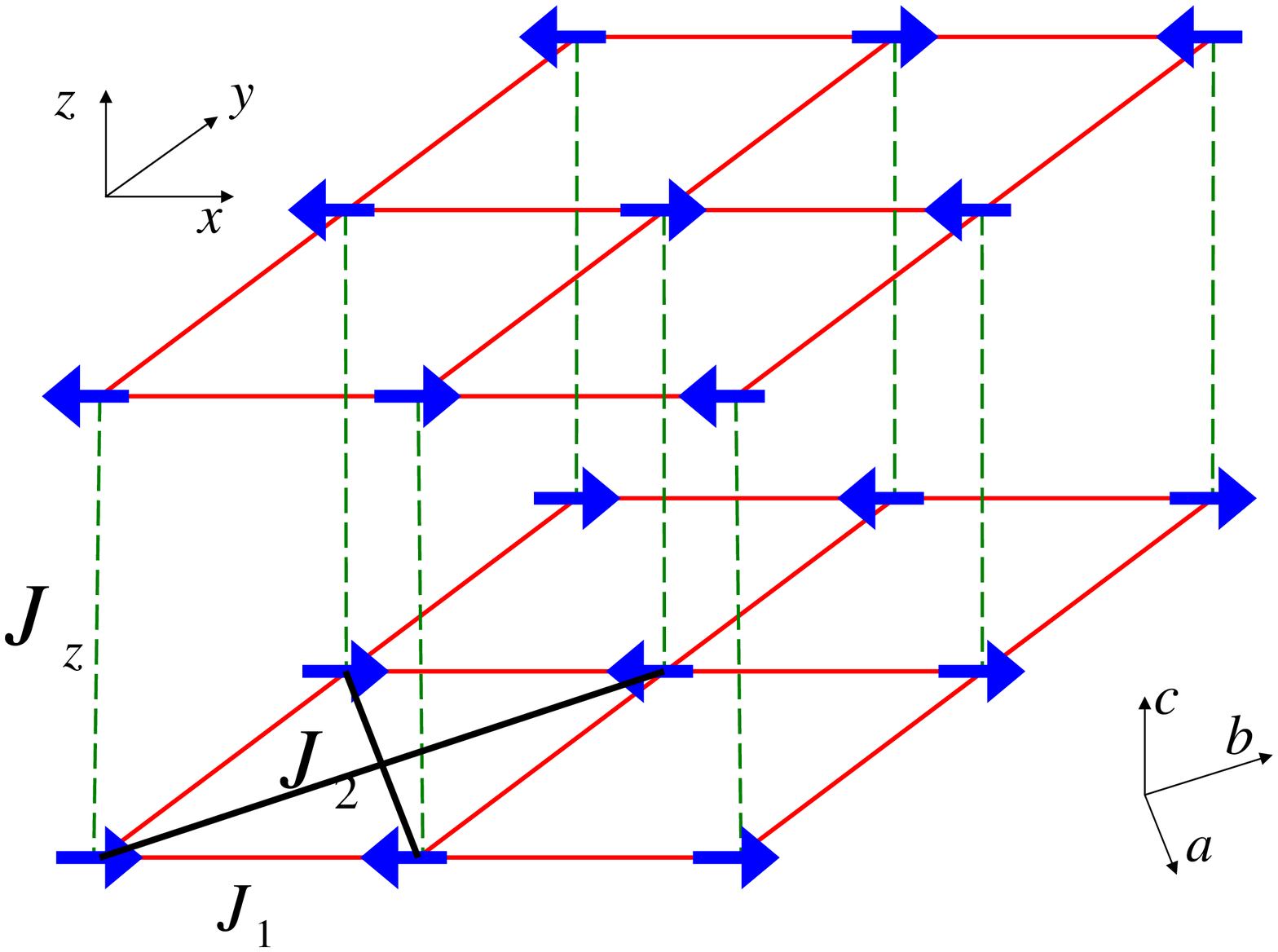}% Here is how to import EPS art
\caption{Schematic graph for the proposed model with nearest-neighbor coupling $J_1$, next-nearest-neighbor coupling $J_2$ and interlayer coupling $J_z$. The orientation of the spins in the low temperature phase are drawn according to Ref. \onlinecite{mook}. Note that we use coordinate system with axis $x$, $y$ $z$ in the current study, which is $45^\circ$ rotated along the $c=z$ direction from the realistic crystal axis $a$, $b$ and $c$. } \label{fig:model}
\end{figure}

  In the tetragonal phase, the iron sites form square planar arrays, such that the sites of adjacent
  planes lie above one another.  Because the superexchange is mediated through off-plane but
   plaquette-centered As atoms, the first and second neighbor antiferromagnetic exchange couplings, $J_1$ and $J_2$,
    are expected to be of roughly the same magnitude. However, the coupling between spins on neighboring planes,
    $J_z$, while still antiferromagnetic, is expected to be much smaller than the in-plane couplings.
    (See Fig.\ref{fig:model}) Estimates from previous work\cite{yildirim,Chandra1990,Xiang} are
     $J_1\approx0.5J_2 \approx 400-700$K. $J_z$ is several orders of magnitude smaller than $J_2$.   The resulting minimal Hamiltonian is
 \ba
 &&H=\sum_{n,{\bf R},{\boldsymbol \delta}_1} \left[J_1\vec S_{{\bf R},n}\cdot \vec S_{{\bf R}+{\boldsymbol \delta}_1,n}
 -K\left(\vec S_{{\bf R},n}\cdot \vec S_{{\bf R}+{\boldsymbol \delta}_1,n}\right)^2\right] \nonumber \\
&&+J_2\sum_{n,{\bf R},{\boldsymbol \delta}_2} \vec S_{{\bf R},n}\cdot \vec S_{{\bf R}+{\boldsymbol \delta}_2,n}
 + J_z\sum_{n,{\bf R}}  \vec S_{{\bf R},n}\cdot \vec S_{{\bf R},n+1} %\nonumber \\
 %+ && J_{\cal N}\sum_{n,{\bf R},{\bf \delta}_1} \left[F_d({\bf \delta}_1)\vec S_{{\bf R},n}\cdot \vec S_{{\bf R}+{\bf \delta}_1,n}\right]
 \label{H}
 \ea
 where $S_{{\bf R},n}$ is a spin $S$ operator on site ${\bf R}$ in plane $n$ and ${\boldsymbol \delta}_1$ and
  ${\boldsymbol  \delta}_2$ are, respectively, the first and second nearest-neighbor lattice vectors in square plane. We set the lattice spacing between nearest neighboring iron sites to be 1. The biquadratic interaction term, $K$, is small for well localized spins, but even if we were to omit this term in
  the bare Hamiltonian, it would rise from quantum or thermal fluctuation through the so called `order out of disorder' mechanism\cite{shender,Henley,Chandra1990,chayes} in the long wavelength limit. Note that this Hamiltonian has the $C_4$ lattice rotational symmetry of the high temperature tetragonal phase. % $F_d$ is a ``d-wave'' form factor,
% $F(\pm{\bf x})=-F_d(\pm{\bf y}) = 1$.
  %In the tetragonal phase, $J_{\cal N}=0$, but

In the broken symmetry ``nematic'' phase, %$\Delta( {\bf x}a) = -\Delta( {\bf y}a)\equiv \Delta$
the spin nematic order parameter
\be
{\cal N}\equiv \langle
\sum_{{\boldsymbol \delta}_1} F_d({\boldsymbol \delta}_1)\vec
S_{{\bf R},n}\cdot\vec S_{{\bf R}+{\boldsymbol
\delta}_1,n}\rangle
\ee
 is nonzero, where $F_d$ is a ``d-wave'' form
factor, $F_d(\pm{\hat {\bf x}})=-F_d(\pm{\hat {\bf y}})=1$. Since a
structural distortion of appropriate symmetry is linearly coupled to the spin nematic order
parameter, the magnitude of the structural distortion $u$
 will be proportional to ${\cal N}$ in the presence of weak electron-lattice coupling. It is a central
conclusion of our work that the structural distortion is a
response to a purely electronic pattern of symmetry breaking, so
we will typically take $u=0$, although in some places, we  will consider the effects of a small
perturbation
\be
H^\prime=J_{\cal N}\sum_{n,{\boldsymbol R},{\bf \delta}_1}
\left[F_d({\boldsymbol \delta}_1)\vec S_{{\bf R},n}\cdot \vec S_{{\bf
R}+{\boldsymbol \delta}_1,n}\right]
\ee
where $J_{\cal N}\propto u$.

There is a subtlety of the crystal structure that is
not apparent in the model presented in Eq. \ref{H}: Because of the
presence of a glide plane, the spins on the Fe sites are all
equivalent, so in the model, there appears to be only one atom per
unit cell. Thus, the distortion which produces a non-zero $u$
looks to be an orthorhombic distortion, in which the elementary
square plaquette become slightly rectangular.  However, because of
the three dimensional placement of the As atoms out of the Fe plane,
there are actually two Fe sites per unit cell, and consequently
the correct classification of the low symmetry phase is
monoclinic. Because it simplifies the discussion, we will
henceforth ignore this subtlety, and phrase our discussion on an
idealized crystal structure with a single Fe atom per unit cell,
as is appropriate to the magnetic Hamiltonian in Eq. \ref{H}.

We will consider this model in the limit $J_2 > J_1/2 \gg
J_z,\ K > 0$. We shall derive $T=0$ properties of this model
to lowest order in a spin-wave ($1/S$) expansion.  To treat the
finite temperature properties of the model, we consider $\vec S$
to be an $N$ dimensional unit vector ($SO(N)$ spin) and obtain a
systematic solution to the problem in the large $N$ limit.  It is
generally found that the spin-wave theory is
accurate\cite{chakravarty} at $T=0$, even in the limit $S=1/2$, so
it should be quite reliable in the present case.  Similarly, the
physical $N=3$ typically is well approximated by large $N$.

\section {Zero temperature (large $S$)}
\label{largeS}

We can compute the properties of the system described in Eq. \ref{H} using the standard (Holstein-Primakov) spin-wave theory.  We compute the classical expression and the leading corrections to order $1/S$.

Since at $T=0$, the small interplane coupling $J_z$ does not qualitatively effect the magnetic properties of the system, we will simplify expressions for various quantities by evaluating them in the limit $J_z \to 0$. For $J_1 > 2J_2 \gg K > 0$, the classical ($S\to\infty$) ground state is the Neel state with ordering wave-vector ${\bf Q}=(\pi,\pi)$.  For $J_2 > J_1/2\gg K >0$, the regime of interest in the present study, the classical ground state is the ``striped'' phase with ordering vector ${\bf Q}=(\pi,0)$, as shown in Fig. \ref{fig:model}, or $(0,\pi)$.  Note that, in addition to breaking spin-rotational symmetry and time reversal symmetry, this state (even in the absence of any spin-orbit coupling) spontaneously breaks the lattice symmetry.  Specifically, we compute the antiferromagnetic and the nematic order parameters,
\ba \label{O}
&&m =e^{i{\bf Q}\cdot {\bf R}}\langle {\vec S}_{{\bf R},n}\rangle= S - \alpha +\ldots  \\
&& {\cal N} = \langle \ \sum_{{\boldsymbol \delta}_1}F_d({\boldsymbol \delta}_1){\vec S}_{{\bf R},n}\cdot{\vec S}_{{\bf R}+\hat {\boldsymbol \delta}_1,n}\ \rangle
 = -4[S^2-\beta S +\ldots]
 \nonumber
\ea
where $\alpha$ and $\beta$ are dimensionless functions of $J_2/J_1$ that we compute below.  Here, we have taken the ordering vector ${\bf Q}$ to be $(\pi,0)$ as in shown Fig. \ref{fig:model}, which means the nearest-neighbor bonds along the x direction are satisfied, but the y-directed bonds are ferromagnetic, and ``frustrated.''

The spin wave spectrum to leading order in $S$ is given by
\ba
&&\omega^2_{{\bf k}}= 4S^2\left[\ \left(J_{y}\cos k_y+2J_2+J_x-J_y\right)^2\right.  \nonumber \\
&&\qquad -\left. \left(J_{x} + 2J_2\cos k_y \right)^2 \cos^2 k_x \right],
\ea
where $J_y\equiv J_1+2J_{\cal N}- 2KS^2$ and $J_x\equiv J_1-2J_{\cal N}+ 2KS^2$.  As expected, the spectrum is gapless at
 ${\bf k}=(0,0)$ and ${\bf k}={\bf Q}=(\pi,0)$.  These are the Goldstone modes, which have an anisotropic linear dispersion,
 \ba
% \omega_{\bf k} \approx 2S \left[(2J_2+J_{x})^2 q_x^2 +(2J_2+J_{x})(2J_2-J_{y})q_y^2\right]^{\frac{1}{2}},\\
 \omega_{\bf k} \approx 2S \sqrt{(2J_2+J_{x})\left[(2J_2+J_{x}) q_x^2 +(2J_2-J_{y})q_y^2\right]},~
\ea
in which ${\bf q}$ is the small deviation from the gapless points.
The spectrum is also almost gapless at ${\bf k}={\bf Q}^\prime=(0,\pi)$ and ${\bf k}={\bf Q}^{\prime\prime}=(\pi,\pi)$, where the gap,
\ba
\Delta=4S\sqrt{(J_{x}-J_{y})(2J_2-J_y)}
\ea
 is determined by the small terms in the Hamiltonian.  The ordered state can be thought of as two inter-penetrating Neel states, which at the classical level do not lock to each other.  It is the small terms, $K$ and $J_{\cal N}$ which lock them together in a collinear state, and gap what would otherwise be two independent sets of Goldstone modes.  Notice that this gap also vanishes at the critical coupling $J_2 \to J_1/2$.
This is somewhat surprising, as the striped to Neel transition might otherwise be expected to be first order. (It is an open question, which we will not address at present, whether there is an interesting, unconventional\cite{deconfined} quantum critical point, here, or possibly some additional intermediate zero temperature phases stabilized by quantum fluctuations.)

At the same level of approximation, we can compute the leading order quantum corrections to the sublattice magnetization and nematic order parameters, $\alpha$ and $\beta$ in Eq. \ref{O}.  To simplify matters, we compute both quantities in the limit %$J_{\perp}$,
$K$ and $J_{\cal N} \to 0$, since these small couplings make only negligible differences in the results.  Then
\ba
&&\alpha=
\frac{1}{2}\Big[\int \frac{d^2\bf k}{(2\pi)^2}\frac{2(J_1\cos k_y+2J_2)}{\omega_{\bf k}/S}-1\Big],
\ea
and
\ba
&&\beta =2\alpha-\int\frac{d^2\bf k}{(2\pi)^2}\frac{J_1(\cos^2 k_x+\cos^2 k_y)}{\omega_{\bf k}/S}\nonumber\\
&&\qquad-\int\frac{d^2\bf k}{(2\pi)^2}\frac{2J_2\cos k_y(1+\cos^2 k_x)}{\omega_{\bf k}/S}.
\ea
 These integrals are readily evaluated numerically.  For instances, for $J_2/J_1=2.0$, $\alpha=0.20$, $\beta=0.30$, and for $J_2/J_1=1.0$, $\alpha=0.22$, $\beta=0.21$. Both $\alpha$ and $\beta$ diverge as $J_2 \to J_1/2$, but only logarithmically, $\alpha,\beta \sim (2\pi)^{-1}\ln[J_1/(2J_2-J_1)]$.  Thus, except extraordinarily close to the quantum critical point, quantum fluctuations do not significantly reduce the ordered moment.
 %, contrary to the speculation of Si and Abrahams\cite{si}.

 The dynamic structure can also be readily computed.  The transverse piece has the form
\be
S_{\perp}({\bf k},\omega) = A({\bf k})\delta(\omega - \omega_{\bf k}),
\ee
where $A({\bf k})=4\pi S^2\big[J_x(1-\cos k_x-J_y(1-\cos k_y)+2J_2(1-\cos k_x \cos k_y)\big]/\omega_{\bf k}$. Note that interesting behavior is observed near ${\bf k} = {\bf Q}^\prime$ and ${\bf k} = {\bf Q}^{\prime\prime}$: $A({\bf Q}^\prime)=%A({\bf Q}^{\prime\prime})=
2\pi S\sqrt{(2J_2-J_y)/(J_x-J_y)}$, which is large; %since $J_x-J_y$ is much smaller than $2J_2-J_y$. 
however, $A({\bf Q}^{\prime\prime})=2\pi S\sqrt{(J_x-J_y)/(2J_2-J_y)}$, which is small. 
%[\sqrt{\frac{J_x-J_y}{2J_2-J_y}} +\sqrt{\frac{2J_2-J_y}{J_x-J_y}}\Big]$ is quite large.
The longitudinal structure factor has the form of a two spin-wave continuum, and can also be computed explicitly.\cite{hong}

\section { Finite $T$ (large $N$) solution}

To study the phase diagram at finite temperature, and in particular to gain insight into the regime of fluctuating magnetic order above the spin ordering transition temperature, it is sufficient to treat the problem classically, as the effects of quantum fluctuations simply produce small renormalizations of the effective parameters, as above.  %To simplify the calculations, we also take a continuum limit.
Since we are interested in the region $J_2> J_1/2$, we break the system up into two interpenetrating square lattices on which $J_2$ is the nearest-neighbor coupling.  On each sublattice we define the staggered magnetization, $\vec \phi_{n,\alpha}$ for plane $n$ and sublattice $\alpha=1$ or 2. % depending on the sublattice.
  To make our calculations analytically tractable, we take the continuum limit and for convenience
  we write the model with respect to the real
crystal axis, i.e. $x,y$ are equivalent to $a,b$ crystal directions respectively in the following model. So \ba
 H_{c}&=& \int d^2{\bf r}\sum_{n,\alpha}\Big[\frac 1 2 \tilde J_2|\nabla {\vec \phi}_{n,\alpha}({\bf r})|^2
-\tilde J_z
\vec\phi_{n,\alpha}({\bf r})\cdot \vec \phi_{n+1,\alpha}({\bf r})\Big]  \nonumber \\
 &&- \tilde K\sum_{n}\left[\vec \phi_{n,1}({\bf r})\cdot {\vec \phi}_{n,2}({\bf r})\right]^2\nonumber\\
&& + \frac{\tilde J_1}{2}\sum_{n}{\vec \phi}_{n,1}({\bf
r})\partial_x\partial_y{\vec\phi}_{n,2}({\bf r}),~~~~~~ \ea where we
have used the same symbols with a tilde for the couplings as in Eq.
\ref{H}, although the present quantities should include
renormalizations due to both quantum effects and high energy thermal
fluctuations. $\tilde J_i\sim J_iS^2, (i=1,2)$ and $\tilde J_{z}\sim
J_z $. Specially, as we pointed out before, the $\tilde K$ term can
be shown\cite{Chandra1990} to rise through fluctuations in the
following form $\tilde K\sim0.13\tilde J_1^2S^2/\tilde J_2$, which
is about $0.01\tilde J_2$ with the estimates of $J_1$ and $J_2$
given in Ref. \onlinecite{yildirim}.

We take the large $N$ limit and solve the problem.  The
self-consistency equations can be derived using the same method
employed in Ref. \onlinecite{Fang2006}.  Define  the nematic order
$\sigma =2\tilde K \langle \vec \phi_{n,1}({\bf r})\cdot\vec
\phi_{n,2}({\bf r})\rangle $ and $\lambda_{n,\alpha}({\bf
r}),\alpha=1,2$ are the Lagrangian multiplies for
$\vec\phi_{n,\alpha}({\bf r})$. The saddle point of above Lagrangian
is determined by  the following self-consistent equations where we
take $\lambda_{n,\alpha}({\bf r})=\lambda, \alpha=1,2$:
\ba
\sigma=\frac{-2\tilde K}{(2\pi)^3}\cdot\int_{-\Lambda}^{\Lambda}dk_x
\int_{-\Lambda}^{\Lambda}dk_y\int_{0}^{2\pi}dk_z G_{12}(\vec k)
\ea
\ba 1=\frac{2T}{(2\pi)^3}\cdot \int_{-\Lambda}^{\Lambda}dk_x
\int_{-\Lambda}^{\Lambda}dk_y\int_{0}^{2\pi}dk_z G_{11}(\vec k) \ea
where $\Lambda\sim {\cal O}(1)$ is momentum cutoff, and
\ba G(\vec k)^{-1}=
 \left(\begin{array}{cc}
A(\vec k) & B(\vec k) \\
B(\vec k) & A(\vec k)
\end{array}\right),
\ea
where $A(\vec k)= {\tilde J}_2 k^2+2\tilde J_z \cos k_z+2\lambda T$,
$B(\vec k) =2T\sigma+ \tilde J_1 k_xk_y$ with $\vec k=({\bf k},k_z)$ and $k^2=k_x^2+k_y^2$.
From these self-consistent equations, we can determine the
transitions temperatures. For ${\tilde K}, \tilde J_z \ll {\tilde
J}_2$, the nematic transition temperature $T_{\cal N}$ is determined
by the following equation
\ba \frac{2\pi {\tilde J}_2}{T_{\cal
N}}=ln\frac{{\tilde J}_2/T_{\cal N}}{\sqrt{(\frac{{\tilde K}}{4\pi
{\tilde J}_2})^2 +(\frac{\tilde J_z}{T_{\cal N}})^2}+\frac{\tilde
K}{4\pi \tilde J_2}}
\ea

\begin{figure}[t]
\includegraphics[width=9cm, height=6cm]{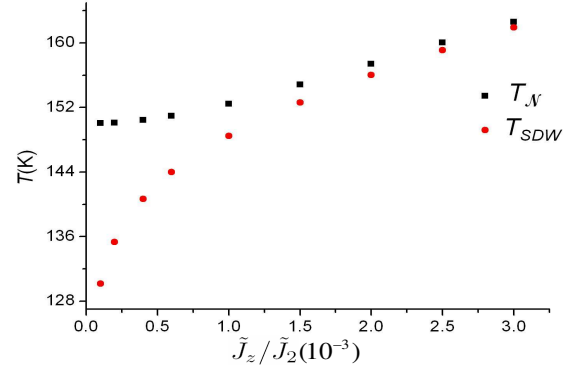}
\caption{$T_{\cal N}$ and $T_{SDW}$ as the function
of $\tilde J_z$ for $\tilde J_2=2\tilde J_1$ and $\tilde K=0.01\tilde J_2$. The realistic value of $\tilde J_z$ can be determined by setting $T_{\cal N}$ around $150K$.}
\label{fig:phase}
\end{figure}

\begin{figure}[t]
\includegraphics[width=9cm, height=6cm]{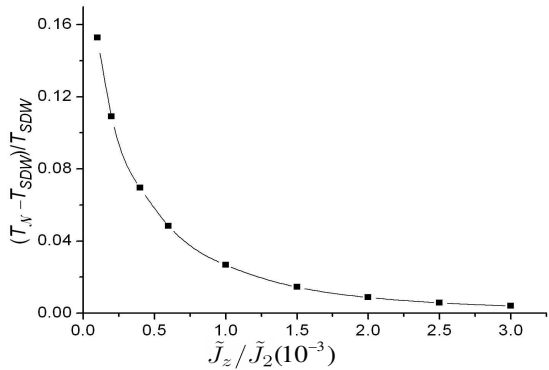}
\caption{$\frac{T_{\cal N}-T_{SDW}}{T_{SDW}}$
as the function of $\tilde J_z$ for  $\tilde K=0.01\tilde J_2$.}
\label{fig:difference}
\end{figure}

The spin density wave transition temperature $T_{SDW}$ takes place
when $\lambda=\sigma_{SDW}+\tilde J_z/T$. It is determined by the
following equations
\ba
\label{neel}
&& \frac{\sigma_{SDW}}{ 2\tilde
K}=\frac{1}{8\pi \tilde J_2} ln\frac{2\sigma_{SDW}+\frac{\tilde
J_z}{T_{SDW}}+2\sqrt{\sigma_{SDW}^2+\frac{\sigma_{SDW}
\tilde J_z}{T_{SDW}}}}{\tilde J_z/T_{SDW}},\nonumber\\
& & \frac{\sigma_{SDW}}{2\tilde K}+\frac{1}{2T_{SDW}}=\frac{1}{4\pi
\tilde J_2}ln\frac{\tilde J_2}{\tilde J_z}.
\ea
By solving these equations, we find that the above model has two
second order phase transitions. The nematic transition temperature
$T_{\cal N}$ is always larger than the SDW transition temperature
$T_{SDW}$. In Fig.\ref{fig:phase}, we show the transition
temperatures $T_{\cal N}$ and $T_{SDW}$ as the function of $\tilde
J_z$  for a fixed $\tilde K =0.01\tilde J_2$. $T_N$ is largely
insensitive to $\tilde J_z$ so long as it is small. In Fig.
\ref{fig:difference}, we plot the difference
$
{(T_{\cal N}-T_{SDW})}/{T_{SDW}}$ as the function of $\tilde
J_z$. If we compare the experimental result in LaOFeAs where
$
{(T_{\cal N}-T_{SDW})}/{T_{SDW}} \sim 11\%$, our results suggest
$\tilde J_z\sim  10^{-4} \tilde J_2$. In this parameter region, by
increasing  $\tilde J_z$, which can be achieved by applying external
pressure along $z$ axis,  the difference between $T_{\cal N}$ and $T_{SDW}$ can be
reduced.

\section{Final Remarks}

The localized spin model we have solved produces results that have enough in common
with the observed ordered phases of ROFeAs that, we believe, it is clear that it captures some of the correct physics, as was first proposed by Yildirim\cite{yildirim} and Si and Abrahams.\cite{si}  Both the stripe-like pattern of magnetic order and the existence of a monoclinic (nematic) lattice distortion are found to be inevitable consequences of the magnetic interactions.  By analyzing the model carefully, we have reached some additional conclusions.

1) Most importantly, we find that there is inevitably a narrow range of temperatures above
the magnetic ordering temperature, in which nematic order persists.  The tendency toward nematic order
lifts the frustration and permits coupling, even ignoring fluctuations, between the staggered magnetic
order on the two sublattices.  Thus, the magnetic order is certainly enhanced by a monoclinic lattice distortion,
as emphasized by Yildirim.  However, we have shown that the symmetry reduction can be purely magnetic
 in origin (with a resulting lattice distortion occuring as a secondary consequence of this ordering),
 and that indeed the symmetry reduction occurs when there is only intermediate scale magnetic order.\cite{chayes}
  In this sense, the nematic phase in RLnFeAs is a consequence of fluctuating magnetic stripe order.
  This observation can be tested by studying the evolution of the fluctuating magnetic order in the
   temperature interval between $T_{SDW}$ and $T_{\cal N}$ using inelastic neutron scattering.

2)  In contrast to a speculation of Si and Abrahams, we find that the frustration inherent
in the model is not sufficient to account for the experimentally observed small magnitude of the ordered moment. Rather, this reflects an intrinsic shortcoming of the strong-coupling model.  Presumably, as the electrons giving rise to the spin become increasingly weakly localized, the maximum possible magnitude of the ordered moment decreases.

3) The validity of the strong coupling approach can be directly tested by looking for additional collective modes that the model predicts.  In particular, in addition to the Goldstone modes associated with the broken spin rotationally symmetry (which are presumably weakly gapped due to spin-orbit coupling), there should be additional almost Goldstone modes at wave-vectors ${\bf Q}^\prime$ and ${\bf Q}^{\prime\prime}$, whose gap is a direct reflection\cite{shender} of quantum fluctuations of the spin through ``order from disorder.''

Finally, we conclude with some more speculative remarks:

1)  Many papers\cite{Kamihara08,Dong,McGuire,mook} have observed a transition temperature, $T_\rho$, at which there is a relatively sharp feature in the resistivity.  While this transition has been widely identified as a spin-density wave transition, in LaOFeAs, for which direct neutron scattering evidence is available, it actually occurs at $T_{\cal N}$ rather than at the 15K lower $T_{SDW}$.  This suggests that $T_\rho = T_{\cal N}$, more generally.  With increasing doping, $x$, $T_\rho$ drops and so, presumably does both $T_{\cal N}$ and $T_{SDW}$.  Following the logic of the present paper, it is probable that $T_{SDW}\to 0$ at a critical doping concentration, $x_{SDW}$ which is less than the critical concentration, $x_{\cal N}$, at which $T_{\cal N} \to 0$.  If we accept the identification between $T_\rho$ and $T_{\cal N}$, then extrapolating the results to where $T_\rho \to 0$, one would conclude that $x_{\cal N}$ is greater than the minimum $x_c$ for superconductivity, and indeed roughly coincides with the point at which the superconducting $T_c$ first reaches its maximum value.  However, contrary to the inference made in Ref. \onlinecite{liu} in which this feature was observed, this does not necessarily imply the coexistence of superconductivity and magnetism;  it is possible that $x_{SDW} \leq x_c < x_{\cal N}$.

2)  Si and Abrahams conjectured that the superconductivity in the oxypnictides may have d-wave symmetry.  While we are still uncertain as to how far the present strong-coupling approach can be extended, it does seem natural from the present perspective that the d-wave character of the electron nematic order could carry over to a d-wave character of the proximate superconducting state.

3)  The existence of a form of stripe order, and of a nematic phase associated with fluctuating stripe order, would constitute a striking piece of evidence of a close relation between the physics of the cuprate and the oxypnictide superconductors.  Stripe order, albeit with a longer period and a different character, has been observed for many years in a variety of cuprate superconductors.\cite{rmp} Fluctuating stripe order has been detected in neutron scattering studies of a much broader class of cuprates.  Moreover, recently, it has been confirmed in neutron scattering studies\cite{keimer,mooknematic} of underdoped YBCO, that a nematic phase associated with fluctuating stripe order occurs at a temperature well above the superconducting $T_c$, which presumably vanishes at a critical doping somewhere inside the superconducting dome.

{\it Acknowledgments} JPH and CF were supported by the National
Science Foundation under grant No. PHY-0603759. SAK was supported in
part by DOE grant No. DE-FG02-06ER46287. HY and WFT were supported in
part by DOE grant No. DE-AC02-76SF00515. HY is also supported by a Stanford SGF.


\begin{thebibliography}{23}
\bibitem{bad} Bad metals are defined\cite{emerykivelsonbad} to be materials with a resistivity, $\rho$, which increases with increasing temperature (as in a ``good metal''), but with a sufficiently large magnitude that, were $\rho$ to be interpreted in terms of quasiparticle transport, it would imply a mean-free path shorter than the Fermi wavelength.  As was pointed out in Ref. \onlinecite{si}, for parameters characteristic of SmOFeAs, taking $\rho = (h/e^2) (k_F l)^{-1} a_c=7$m$\Omega$-cm and a spacing between Fe planes of $a_c = 8.7\AA$, we would conclude that $l \sim 0.1 \lambda_F$ where $\hbar k_F = h/\lambda_F$ is the Fermi momentum.
\bibitem{caveat}
Since resistivity measurements to date are carried out on polycrystalline samples, it may be that the measured resistivities are larger than the intrinsic resistivity;  when resistivity data becomes available on single crystals, the classification of ROFeAs as a bad metal may need to be revisited.
\end{thebibliography}
\end{document}